\documentclass[12pt, preprint]{aastex}
\usepackage{amsmath, amsthm}
\input epsf

\newcommand{\partt}{\frac{\partial}{\partial t}}
\newcommand{\partR}{\frac{\partial}{\partial R}}
\newcommand{\di}{\vec{\nabla} \cdot}
\newcommand{\grad}{\vec{\nabla}}
\newcommand{\lb}{\langle}
\newcommand{\rb}{\rangle}
\newcommand{\ti}[1]{\tilde{#1}}
\newcommand{\tb}[1]{\textbf{#1}}
\newcommand{\ov}{\overline}
\newcommand{\fl}{\text{fluc}}

\begin{document}
\title{Identifying Deficiencies of 
Standard Accretion Disk Theory: Lessons from a Mean-Field Approach}
\author{A. Hubbard\altaffilmark{1,2}, E.G. Blackman\altaffilmark{1,2}}
\affil{1. Dept. of Physics and Astronomy, Univ. of Rochester,
    Rochester, NY 14627; 2. Laboratory for Laser Energetics Univ. of Rochester, Rochester NY, 14623}

\begin{abstract}
Turbulent viscosity is frequently used in accretion disk theory to replace
the microphysical viscosity in order to accomodate the observational
need for instabilities in disks that lead to
enhanced transport. However, simply replacing  the microphysical
transport coefficient  by a single turbulent  transport coefficient 
hides the fact that the procedure should formally arise
 as part of a closure in which the hydrodynamic or magnetohydrodynamic 
equations are averaged, and correlations of turbulent fluctuations 
are replaced by transport coefficients. 
 Here we show how a mean field approach  leads quite naturally two transport coefficients, not one,  that govern 
mass and angular momentum transport. In particular, 
we highlight that the conventional
approach suffers from a seemingly inconsistent neglect of turbulent diffusion
 in the surface density equation.  We constrain these new transport coefficients  for
specific cases of  inward, outward, and zero net mass transport. 
In addition, we find that 
one of the new transport terms can lead to oscillations in the mean
surface density which then requires a constant or small inverse Rossby number
for disks to maintain a monotonic power-law surface density.

\end{abstract}


\section{Introduction}

Accretion disks are ubiquitous around stars and compact objects \citep{kingfrankraine}.
Those around black holes produce some of the most luminous objects in the universe, while those around young stars provide the material from which planets
form. Thin astrophysical accretion disks are primarily 
angular momentum supported and thus to be accreting, 
require a mechanism to export angular momentum.  Observations  suggest 
that  micro-physical viscosity acting on the differential rotation in a near-Keplerian disk is too weak to accommodate needed time scales and luminosities, given
mass density constraints \citep{pringle1981} 
and so  attention has focused on turbulence
to enhance the needed transport. While outflows are also 
likely important, the focus on turbulent transport raises three issues:
(1) What are the minimum properties that turbulence must have to
transport angular momentum outward? (2) What causes the turbulence?
 (3) What 
 practical set of equations  captures the correct physics?

Turbulence was first incorporated 
into a practical analytic formalism by Shakura-Sunyaev \citep{shak}.
There, the only role of turbulence emerges as a simple replacement
of the micro-physical viscosity in the momentum equation with an enhanced
turbulent viscosity. One way that this immediately falls short is that 
the current leading candidate for turbulent transport is the magneto-rotational
instability (MRI) which involves magnetic fields not considered
in Shakura-Sunyaev formalism.
That being said, the MRI drives turbulence and one might hope
that  the resulting transport could be modeled via a Shakura-Sunyaev
formalism.  Evidence is emerging that this is not the case
(\citealt{pessah2}, \citealt{pessah3}, \citealt{blackmanpenna})
and there presently remains a disconnect between the insights gained
from numerical simulations and a practical formalism that can both 
accommodate the MRI physics and be of value to phenomenological modelers
(e.g. \citealt{pessah2}, \citealt{ogilvie03}).

As part of a long term effort 
to develop a formalism which captures 
the minimum properties that turbulence must have 
to transport angular momentum outwards  it is instructive to 
take a step back from the MRI and rethink the basic meaning of
accretion disk equations.
Any axisymmetric model that includes turbulent transport is implicitly
a mean field theory in that locally, turbulence breaks axisymmetry.
Therefore equations that evolve quantities presumed to be 
only a function of radius $R$ are only relevant if they can be derived
from a plausible (even if crude) mean field formalism.

Here we revisit the hydrodynamic accretion disk equations
 from a mean field point of view and 
evaluate whether the standard presence of turbulent transport as 
merely an enhanced viscosity emerges from  plausible assumptions from
a mean field theory. We find that it does not. Instead we find 
transport terms in the surface equation and in the angular momentum
transport equation that are  missing  from the standard $\alpha$ viscosity
formalism. This  leads to  conditions on the properties of these transport
terms that allow outward transport of angular momentum.
That only turbulence with specific properties can transport 
 angular momentum outward is suggested by the simple fact that 
 specific angular momentum in a Keplerian disk 
increases with radius and therefore one might initially expect 
turbulent diffusion to transport angular momentum \textit{inwards}.  In at least some systems this expectation is confirmed, as studies of convective turbulence show this effect (\citealt{R&G1992}, \citealt{K1993}) and indeed one of the virtues of MRI is that it is expected to drive accretion \citep{B&Hreview}.

In Sec. 2 we derive our mean field formalism and derive
the evolution equations for the mean surface density and mean orbital
angular momentum.  We compare the results to the conventional
versions of these equations in Sec. 3. In Sec. 4 we discuss
the implications of the new equations for outward angular momentum
transport and steady state disks. We conclude in Sec. 5.

\section{Mean Field Formalism}

We   distinguish three relevant time scales for the accretion disk: (1)
a long global time-scale $t_g$ corresponding to accretion of the entire disk; (2) a  dynamical time-scale $t_d \le t_g$ corresponding to the local accretion time-scale; (3) a short time-scale $\tau \ll t_d$ corresponding to the energy containing eddy turnover time.  We assume that that the disk is thin (height $h\ll R$) and that the time averaged disk is equivalent to an axially symmetric and
vertically integrated disk.  While full vertical averaging is appropriate for extracting fluxes, some physics is lost (e.g. \citealt{Blackman01}).  Accordingly, we  adopt cylindrical coordinates $(R, \phi, z)$.   We assume that the largest turbulent eddies are small compared to the radial scale at all radii, that is, $\lambda \ll R$, where $\lambda$ is the
eddy scale. 
We then decompose all quantities into fluctuating and mean values
such that  $q={\overline q} + {\tilde q}$ where  $q$ is a flow quantity
where $\overline q=\lb q \rb$ represents the mean and $\tilde q$ is the fluctuation about the mean. The fluctuations 
 average to zero over time-scales longer than $\tau$ 
and the means have  temporal variation scales much larger than  $\tau$ \citep{moffat}.

In addition, we  make the important simplifying anzatz that
fluctuating quantities can be separated into a factor  that varies on 
spatial scale $R$ times an isotropic homogeneous  fluctuation.
For the surface density and the velocity in particular, we write
\begin{equation}
\ti{\Sigma} = \Sigma_{\ast} {u}_{\Sigma}
\ \ {\rm and} 
\ \ \ti{\tb{v}}=v_{\ast} \tb{u}_{v} 
\label{1}
\end{equation}
where 
$ v_* \equiv\lb\ti{\tb{v}}^2 \rb^{1/2}$, 
 $\tb{u}_{v}$  is a statistically isotropic and homogeneous vector, $\Sigma_* \equiv \lb \ti{\Sigma}^2 \rb^{1/2}$ and  ${u}_{\Sigma}$  is a statistically  homogeneous scalar.
This decomposition is similar to a WKB approximation and will
prove useful in section 3.

  Using the above decomposition, we now 
produce a mean-field derivation of the evolution of a conserved quantity with areal density $q$, the relevant 
examples of which are surface density $\Sigma$ 
or orbital angular momentum density $L_z$. For such a conserved quantity, we have
\begin{equation}
\partt q+\di (q\tb{v})=0=\partt \ov{q}+\partt \ti{q}+\di(\ov{q}\ov{\tb{v}}+\ti{q} \ov{\tb{v}}+\ov{q} \ti{\tb{v}}+\ti{q} \ti{\tb{v}}). \label{fullexp}
\end{equation}
Taking the time-average (as described above) of Eq. (\ref{fullexp}) we find
\begin{equation}
\partt \ov{q}+\di(\ov{q}\ov{\tb{v}})+\di \lb \ti{q} \ti{\tb{v}} \rb=0, \label{timeav}
\end{equation}
where transport terms akin to diffusion or viscosity derive from the term $\di \lb \ti{q}\ti{\tb{v}} \rb$.
Subtracting Eq. (\ref{timeav}) from Eq. (\ref{fullexp}) we find:
\begin{equation}
\partt \ti{q}=-\di(\ti{q} \overline{\tb{v}}+\ov{q}\ti{\tb{v}}+\ti{q} \ti{\tb{v}}-\lb \ti{q}\ti{\tb{v}} \rb). \label{partq'}
\end{equation}
We approximate the implications of Eq. (\ref{partq'}) at a point $p$ by estimating the time integrated fluctuating flux of $q$ through the surface $S$ composed of those points a distance $\lambda=v_{\star} \tau$ from the point $p$ and approximating it as constant for a time $\tau$.  The distance $\lambda$ is determined at the surface $S$ and is chosen because it encloses the volume $V$ that the point $p$ can exchange material with over a time $\tau$.  This allows us to write Eq. (\ref{partq'}) as:
\begin{equation}
\ti{q} \simeq -\frac{\tau}{V} \int_S \underbrace{(\ti{q} \overline{\tb{v}}+\ov{q}\ti{\tb{v}}+\ti{q} \ti{\tb{v}}-\lb \ti{q}\ti{\tb{v}} \rb)}_{\vec{F}_q} \cdot d\vec{S} \simeq -\frac{\tau \lambda}{V} \di \lambda^{n-1}(\ti{q} \overline{\tb{v}}+\ov{q}\ti{\tb{v}}+\ti{q} \ti{\tb{v}}-\lb \ti{q}\ti{\tb{v}} \rb). \label{centering}
\end{equation}
In the above Eq. (\ref{centering}) $\vec{F}_q$ is the fluctuating flux of $q$ as given in Eq. (\ref{partq'}) and $n$ is the number of spatial dimensions, with $n=2$ for our thin disk.  The factor $\lambda^{n-1}$ in the right-most term in Eq. (\ref{centering}) comes from the fact that under our definitions, $S \propto \lambda^{n-1}$.  In producing Eq (\ref{partq'}), our use of a correlation time $\tau$ has therefore lead us to construct an effective divergence given by $\di_{\fl} \equiv \lambda^{-(n-1)} \di \lambda^{n-1}$ that is centered on $p$ not in units of distance, but rather in units of correlation lengths.  Note that this generalized divergence reduces to the basic divergence only if $\lambda$ is independent of position.  In the case of an accretion disk, $\lambda$ does depend on position and varies on the scale $R$.  Upon applying $n=2$ to Eq. (\ref{centering}) and using $\lambda=v_* \tau$ we then obtain:
\begin{equation}
\ti{q}\simeq-\frac{\tau}{\lambda} \di \lambda(\ti{q}\ov{\tb{v}}+\ov{q}\ti{\tb{v}}+\ti{q}\ti{\tb{v}}-\lb \ti{q}\ti{\tb{v}} \rb ) = -\frac{1}{v_{\ast}} \di \tau v_{\ast}(\ti{q}\ov{\tb{v}}+\ov{q}\ti{\tb{v}}+\ti{q}\ti{\tb{v}}-\lb \ti{q}\ti{\tb{v}} \rb )\label{q'1}.
\end{equation}  
We can use Eq. (\ref{q'1}) to find
\begin{equation}
\di \lb \ti{q} \ti{\tb{v}} \rb=-\di \lb \tb{u}_v 
\underbrace{\di}_{\text{inner }\di} 
[\tau v_{\ast}
(\underbrace{\ti{q}\ov{\tb{v}}}_{(i)}+\underbrace{\ti{q} \ti{\tb{v}}}_{(ii)}+\underbrace{\ov{q} \ti{\tb{v}}}_{(iii)})]\rb
\label{divcor}
\end{equation}
where we have not written the term 
$\di \lb \tb{u}_v {\di} 
{\lb \ti{q} \ti{\tb{v}} \rb} \rb $
which vanishes
as it is the average of fluctuation times a mean.
We now discuss how (\ref{divcor}) simplifies
 for our quantities of interest: surface mass density $\Sigma$ and angular momentum surface density $L=L_z=\Sigma R^2 \Omega$.

For the case  $q=\Sigma$, we first note that 
 $\Sigma_{\ast}\ll \ov{\Sigma}$ because turbulent density fluctuations
do not dominate for a non-self gravitating disk with  
$\tau\ll t_d$, and accretion disks have inner boundaries of no return.  More specifically, in a  near-steady state  for which $t_d\ll t_g$, Eq. (\ref{timeav}) implies that 
\begin{equation}
\di(\ov{\Sigma}\ov{\tb{v}}) \sim \di \lb \ti{\Sigma} \ti{\tb{v}} \rb.
\label{mc}
\end{equation}
  Only the radial derivatives contribute so only the radial
mean velocity contributes to the divergence.
and thus
 $\ov{v}_R/v_{\ast} \sim \Sigma_{\ast} / \ov{\Sigma}\ll 1$. 
This implies term $(i)$ $<<$ term $(ii)$ 
and also that both can be dropped, as they are small with respect to term $(iii)$ in (\ref{divcor})
when $q=\Sigma$. Then, upon using the chain rule for the inner derivative
on term $(iii)$ 
 the contribution that comes from the derivative operating on the fluctuating 
$\tilde {\bf v}$ vanishes because it involves a vector correlation
of fluctuating quantities (products of ${\bf u}$ vectors)
which vanish in our decomposition.  
Only the contributions to term $(iii)$ with the
derivative operating on $\ov \Sigma$ contributes.
Finally then, we can write:
\begin{equation}
\di \lb \ti{\Sigma} \ti{\tb{v}} \rb \simeq -\di \lb \grad \tau v_{\ast}^2 \ov{\Sigma} \rb \simeq -\nabla^2(\nu_1 \ov{\Sigma}), 
\label{diff}
\end{equation}
where $\nu_1 \sim \tau v_{\ast}^2$ is the diffusion coefficient.  The diffusion coefficient $\nu_1$ is due to the correlations of fluctuating values.  


We now consider the case $q=L=\Sigma R^2 \Omega$ in (\ref{divcor}).
First we note that term $(i)$ can again be neglected with respect
to term $(ii)$ as this does not depend on the choice of $q$. 
To foster evaluation of the contributions from $(ii)$ and $(iii)$ we note that 
the mean and fluctuating values of $L$ are  given by 
\begin{equation}
\overline{L}= \lb\Sigma \Omega R^2\rb= {\ov\Sigma} {\ov \Omega} R^2+
\lb {\tilde \Sigma}{\tilde v}_\phi R\rb\simeq
{\ov\Sigma} {\ov \Omega} R^2
\label{meanL}
\end{equation}
and
\begin{equation}
\ti{L}=
\underbrace{\ti{\Sigma}R^2\Omega}_{(a)}+\underbrace{\ov{\Sigma}R\ti{v}_{\phi}}_{(b)}+\underbrace{R(\ti{\Sigma}\ti{v}_{\phi}-\lb \ti{\Sigma} \ti{v}_{\phi}\rb)}_{(c)} \simeq {\ti{\Sigma}R^2\Omega}+{\ov{\Sigma}R\ti{v}_{\phi}}
\label{tiL1}
\end{equation}
respectively.
Note that the fourth term in (\ref{meanL}) is smaller than
the third term for the same reason that 
in (\ref{tiL1}) term   $(c)\ll (a)$ 
and term $(b)\ll (a)$, 
 namely because  $\ti{v}\ll R\Omega$ (thin disk) and  $\ti{\Sigma} \ll \ov{\Sigma}$ as noted in the previous paragraph.  
Then, by analogy to Eq. (\ref{diff}), term $(iii)$ of Eq. (\ref{divcor}) becomes
\begin{equation}
-\nabla^2(\nu_1 \ov L) 
\label{divcor3}.
\end{equation}
Because the transport coefficient deriving from term $(iii)$ of Eq. (\ref{divcor}) is independent of the transported mean quantity $\ov{q}$, 
it is plausible to conclude that the transport coefficients of Eqs (\ref{diff}) and (\ref{divcor3}) are the same, explaining why we have used $\nu_1$ in both equations. 

Unlike the case of surface density, term $(ii)$ of Eq. (\ref{divcor}) can be significant for angular momentum.  Inserting the fluctuating angular momentum density (Eq. \ref{tiL1}) into term $(ii)$ of Eq. (\ref{divcor}) we find
\begin{equation}
-\di \lb \tb{u}_v \underbrace{\di}_{\text{inner } \di} \tau v_{\ast} (\underbrace{\ti{\Sigma} R^2 \Omega}_{(a)}+\underbrace{\ov{\Sigma} R \ti{v}_{\phi}}_{(b)} ) \ti{\tb{v}} \rb \label{tiL2}.
\end{equation}
The dependence of term $(b)$ of Eq. (\ref{tiL2}) on $\ti{v}_{\phi}$ provides a preferred direction, so the averaging  does not eliminate 
the term in which the inner divergence operates on 
 ${\tilde v}_\phi $ in $(b)$.  In fact this is the  dominant contribution from
 $(b)$ since fluctuating quantities vary on scale $\lambda \ll R$.
Using $\vec{\nabla} \sim {1\over \tau v_{\ast}}$, we can write this
contribution from term (b) as
\begin{equation}
\di \lb \tb{u}_v \mu \tau v_{\ast}^2 \ov{\Sigma} v_{\ast}R \frac{1}{v_{\ast} \tau}\rb, 
\label{term2L}
\end{equation}
where we have defined $\mu\equiv
 {\tilde v}_\phi/{\tilde v}_* $
which parameterizes the  dependence on the angular distribution of the velocity fluctuations.  The significance of the parameter $\mu$ can be seen by contemplating the difference between radial and azimuthal turbulent forcing: in a Keplerian disk radial forcing will move material to positions where it has below Keplerian angular momentum while azimuthal forcing, combined with orbital mechanics, will move material to positions where it has above Keplerian angular momentum.
The averaged quantity in Eq. (\ref{term2L}), like all mean values, depends only on $R$ so  only the radial component of  $\tb{u}_v$ survives.  However,
$\mu$ depends on $\ti{v}_{\phi}$. Since 
 orbital motion converts $\ti{v}_{\phi}$ into a radial velocity on a time-scale $\Omega^{-1}$,  only a fraction 
$\sim \tau\Omega \hat{\bf R}$ of $\tb{u}_v$ contributes to the correlation.
Note that $\Omega \tau$ is the inverse Rossby number for the disk.
 Using this in 
(\ref{term2L}), term $(b)$ of Eq. (\ref{tiL2}) can therefore be written 
\begin{equation}
\di \lb \mu \tau v_{\ast}^2 \ov{\Sigma} v_{\ast} R \frac{1}{v_{\ast} \tau} {\tau}{\Omega} \hat{\bf R}\rb =\di \left({\nu_2{\ov{L}}\over R} \hat{\bf R}\right), \label{Lb2}
\end{equation}
where $\nu_2 \sim \tau v_{\ast}^2$ and includes the system's dependence on the angular distribution of the velocity fluctuations (through the dependence of $\nu_2$ on $\mu$, which is not linear because of the averaging process).  
The quantity $\nu_2$ can become negative if the velocity fluctuations along the orbital motion correlate sufficiently strongly with inwards fluctuating motion.  Further, while we expect $|\nu_1| \sim |\nu_2|$, we do not expect $\nu_1=\nu_2$.  In \S 3 we will explore the consequences of different relative values of $\nu_1$ and $\nu_2$.

We have yet to evaluate term $(a)$ of (\ref{tiL2}), but we now argue that
it is much smaller than term $(b)$, the latter contributing Eq. (\ref{Lb2}) to our eventual conservation equations.  We need only compare $\di \ti{\Sigma} R^2 \Omega \ti{\tb{v}}$, (a) to $\di \ov{\Sigma} R \ti{v}_{\phi} \ti{\tb{v}}$, (b).  We note from (\ref{1}) that the divergence on $(a)$ contributes only a surviving term that pulls out a factor of $\sim 1/R$.  Further, we estimate that $\ov{v} \sim {\nu_1\over R} \sim v_{\ast} {\lambda \over R}$.  We can then use Eq. (\ref{mc}) to find $\ti{\Sigma} \sim \ov{\Sigma} \ov{v}/v_{\star} \sim \ov{\Sigma} \lambda/R$.  We find for (a):
\begin{equation}
\frac{\ti{\Sigma} R^2 \Omega}{R} \sim \frac{\ov{\Sigma} \ov{v}  R \Omega}{v_{\ast}} \sim \frac{\ov{\Sigma} \lambda \tau R \Omega}{R} \sim \ov{\Sigma} \lambda \Omega.
\label{15}
\end{equation}
To estimate $(b)$ we note that the inner divergence pulls out a factor ${\Omega \tau \over v_{\ast} \tau}$ where the top comes from the
 azimuthal to radial velocity conversion factor and the denominator comes from the
scale over which an azimuthal turbulent speed varies; note here that this term
results from $v_\phi$ which picks out a preferred direction violating isotropy
and keeping the spatial scale of variation $v_*\tau$ rather than $R$ as in term $(a)$. We thus  have for term $(b)$ 
\begin{equation}
\frac{\ov{\Sigma} R v_{\ast} \tau \Omega}{v_{\ast} \tau} \sim \ov{\Sigma} R \Omega.
\label{16}
\end{equation}
Since $\lambda \ll R$, (\ref{16}) dominates (\ref{15}) so that 
 term $(b)$  dominates term $(a)$ of 
Eq. (\ref{tiL2}).  

We can now finally 
combine Eqs (\ref{divcor3}) and (\ref{Lb2}) 
to find   that
Eq. (\ref{divcor}) with $q=L$
gives
\begin{equation}
\di \lb \ti{L} \ti{\tb{v}} \rb \simeq -\nabla^2(\nu_1 \ov{L})+\di \left({\nu_2 \ov{L}\over R} \hat{\bf R}\right), \label{diffL}
\end{equation}
where  only radial derivatives are non-vanishing.

\section{Conservation Equations and Comparison to Conventional Form}

We can use Eqs (\ref{diff}) and (\ref{diffL}) to write the full, mean-field conservation equations for mean surface density $\ov \Sigma$ and orbital angular momentum surface density ${\ov L}={\ov \Sigma} R^2 \Omega$:
\begin{align}
&\partt {\ov \Sigma}
+\frac 1R \partR( R {\ov v}_R 
{\ov \Sigma})
=\frac 1R\partR \left(R\partR (\nu_1 {\ov \Sigma})\right) \label{consm} \\
&\partt {\ov L}+\frac 1R \partR (R {\ov v}_R  {\ov L})=
\frac 1 R \partR
\left(
R\partR (\nu_1 {\ov L})
-\nu_2 {\ov L}\right). \label{consL}
\end{align}
  The mass flow can be characterized as a turbulent flux 
(right side of Eq. \ref{consm}) 
and a fall-back flux due to orbital angular momentum mismatches (${\ov v}_R$).
For $\Omega \propto R^{-3/2}$ and $\partial \Omega/\partial t=0$, we eliminate
${\ov v}_R$ by combining the above two equations to find the net outwards mass flux
\begin{equation}
F_{M,net}\equiv \partR (\nu_1 {\ov \Sigma})    - {\ov v}_R{\ov \Sigma}=
\partR (\nu_1 {\ov \Sigma}) +\frac{\nu_1 {\ov \Sigma}}{2R}-\frac{2}{R^2\Omega}\partR (\nu_2 {\ov L}). \label{fluxm}
\end{equation}
A finite $\nu_2>0$ is  important in  reducing angular momentum mismatches between
turbulently transported material  and the local material
at the new position.

 Eqs. (\ref{consm}-\ref{fluxm}) 
 can be contrasted to the
analogous equations 
presented in  standard formulations for $\Omega \propto R^{-3/2}$
\citep{kingfrankraine}:
\begin{equation}
\partt {\ov \Sigma} +\frac 1R \partR (R {\overline v}_R
{\ov \Sigma})=0 
\label{angwrong1}
\end{equation}

\begin{equation}
\partt{\ov L}+\frac 1R \partR (R {\overline v}_R{\ov L}) 
=\frac 1 R\partR \left(\frac{\nu  {\ov L}R}{\Omega}\frac{\partial \Omega}{\partial R} \right). 
\label{angwrong2}
\end{equation}
and
\begin{equation}
F_{M} \simeq -{3\nu \ov \Sigma\over 2R}
\label{fluxw}
\end{equation}
where $\nu$ is the Shakura-Sunyaev turbulent viscosity.
These equations differ from 
(\ref{consm}-\ref{fluxm})
  in that there is no turbulent
transport term on the right of 
 (\ref{consm}) and there is only one transport coefficient $\nu$
in the right of (\ref{angwrong2}) as compared to the two transport coefficients
in (\ref{consL}).  
The transport term in the surface density term (\ref{consm})
is particularly noteworthy. The presence of this term highlights that 
the mean radial velocity is not the only contributor to mass motion;
there is also a turbulent diffusion of mass. This is general
expected in a turbulent flow and so its absence in the standard
formalism raises concern.

What is the origin of the differences between 
(\ref{consm}-\ref{fluxm})
 and 
(\ref{angwrong1}-\ref{fluxw}) just presented? 
The usual derivation of the conservation equations {\it does not}
naturally follow from a  mean field derivation as we have shown 
by our derivation of (\ref{consm}-\ref{fluxm}). In particular, the 
standard approach to obtain (\ref{angwrong2})
is to simply replace the micro-physical viscosity in the Navier-Stokes
equation with a turbulent viscosity, assume axisymmetry and vertical integration
 and then derive the angular momentum conservation equation from the momentum
density equation.  Doing so does not produce any transport term in the
surface density equation.  A key point is that although we have written
over-lined quantites $\ov \Sigma$ and $\ov L$ in (\ref{angwrong1}-\ref{fluxw})
we do so only because it is only the mean quantities that have the
axisymmetry and vertical independence usually assumed; 
formally a purely radial dependence   {\it requires} a mean field
formalism  We have therefore have identified an inconsistency with the standard
accretion disk formalism.

Another expression of  the inconsistency is to suppose that a replacement
of micro-physical viscosity with a turbulent viscosity were a legitimate
closure such that no other transport terms appeared and the equations
remained the same.  This would imply that the mean field 
Navier-Stokes equation and turbulent viscosity 
 could be derived  via integration of suitable
 Boltzmann equation 
by analogy to the derivation of the standard Navier-Stokes equation and micro-physical viscosity. However, the derivation of the usual
Navier-Stokes equations comes from the truncation of an expansion in 
the mean free path divided by the macroscopic gradient scale, or a collision
time divided by a macroscopic evolution time.  For accretion disk turbulence,
the latter ratio would be replaced by $\tau\Omega$, a quantity not 
guaranteed to be small, and frequently assumed to be of order unity.
This highlights why a simple replacement of the micro-physical viscosity
by a turbulent viscosity is at best, incomplete.

\section{Conditions for Outward Angular Momentum Transport and Steady-State}

Here we investigate some consequences of 
(\ref{consm}-\ref{fluxm}).
The equation for the net mass flux, Eq. (\ref{fluxm}), can be rewritten as
\begin{equation}
F_{M,net}=\frac{1}{\sqrt{R}}\partR (\sqrt{R} \nu_1 {\ov \Sigma})-\frac{2}{R^2\Omega}\partR (\nu_2 {\ov \Sigma} R^2\Omega). \label{fluxm2}
\end{equation}
This form allows us to explore the consequences of different 
transport coefficients $\nu_1$ and $\nu_2$ and different  radial density profiles.  

We use the standard formula for the disk density scale height 
$h=c_s/\Omega$, where $c_s$ is the sound speed.  We scale our transport coefficients to $c_s h$ as in a Shakura-Sunyaev $\alpha$ disk  prescription \citep{shak}, $\nu_i\sim
\lambda_t v_t=\alpha_i c_s h$, where the $i$ refers to $1,2$ of the previous section.   In a flared disk whose temperature profile is determined by stellar heating ($T \propto R^{-1/2}$), we have $c_s \propto R^{-1/4}$ and $h \propto R^{5/4}$.
If all $\alpha_i$ are constant, then $\nu_i\propto R$.  From Eq. (\ref{fluxm2}) we see  that ${\ov \Sigma} \propto R^{-3/2}$ then results in no net mass transport (although the turbulent and 
fall-back (${\ov v}_R$)  terms of e.g. (\ref{fluxm}) will in general be separately non-zero), and deviations from $ \log_R {\ov \Sigma}=-3/2$ will result in differently signed mass fluxes ($\log_R$ being the base $R$ logarithm).  For a  non-zero steady state mass flux with $\nu_1 \propto \nu_2 \propto c_s h$,  Eq. (\ref{fluxm2}) would imply $\log_R(\nu_i{\ov \Sigma})=0$ .  In that case, the mass flow (positive being outward) becomes
\begin{equation}
\dot{M}=2\pi R F_{M, net}= 2\pi\left( \frac{\nu_1 {\ov \Sigma}}{2}-\nu_2 {\ov \Sigma} \right)=2\pi \left(\frac{\nu_1}{2}-\nu_2\right){\ov \Sigma}, \label{fluxm3}
\end{equation}
and we need $\nu_2>\nu_1/2>0 $ to generate an inwards mass-flux.  Finally, writing $\nu_i \propto v_t \lambda_t$, $\lambda_t =v_t \tau$ and requiring that $v_t \propto c_s$ the condition $\log_R(\nu {\ov \Sigma})=0$ becomes $\log_R({\ov \Sigma} \tau \Omega)=\log_R(\Omega/c_s^2)$.  For the aforementioned flared disk, $\log_R(\Omega/c_s^2)=-1$.

We find another implication  by noting that $\nu_2 \neq 0$ 
implies 
the presence of orbital 
oscillations from (\ref{consL}).  As long as the amplitudes are small, the oscillations can be treated as simple harmonic oscillators and we therefore expect 
oscillating quantities such as ${\ov \Sigma} (R)$ 
to be proportional to $\sin(\tau \Omega)$.  
For  ${\ov \Sigma}(R)$ 
to vary as a power law in $R$, either a constant inverse Rossby number 
$\tau \Omega$ (eddy life time scaling with orbit time) or a small $\tau\Omega$ such that $\sin(\tau \Omega) \sim \tau\Omega$ is then required. 
A short eddy lifetime is interesting as it lowers $\nu$ for a given $v_t$.
(We note that $\tau\Omega >1$ need not be considered because any eddies
with this property initially will be sheared such that
 such that $\tau\rightarrow\Omega^{-1}$; shear makes an otherwise
long lived eddy shred on a rotation time).

\section{Conclusion}

Any turbulent accretion disk model  in which axisymmetry in surface density
is assumed is unavoidably a mean field model and therefore the equations
governing such a model should formally be derived from a mean field theory
with a plausible closure.  
We have presented an explicit mean field approach for deriving
the mean surface density and orbital angular momentum transport
in accretion disks and found that two transport coefficients rather than
just one emerge most naturally. Notably, 
we find  that  the conventional $\alpha$ formalism 
misses a turbulent diffusion  term in the surface density
equation.

In constraining the new transport coefficients  for
specific cases of  inward, outward, and zero net mass transport.
we find that: 
(1) $\log_R(\nu \Sigma)=0$ and $\nu_2 \neq \nu_1/2$ is the condition for a steady-state disk with non-zero net mass transport with 
 $\nu_2>\nu_1/2$  the condition for inwards net mass transport and 
(2) $\log_R(\nu \Sigma)=-1/2$ or $\nu_2=\nu_1/2$ are the conditions for a steady-state disk with zero net mass-transport; 
(3) For a finite $\nu_2$, the inverse Rossby number
$\tau\Omega$ must be  either constant or small for disks to have monotonic
power-law radial dependences of $\ov \Sigma$. This emerges
because  a finite $\nu_2$
implies oscillations in the angular
momentum equation. Such a condition on $\tau\Omega$ 
is not unexpected if one considers that these oscillations depend on 
$\sin (\tau\Omega)$ and since $\tau\Omega \sim 1$ 
blurs the distinction between fluctuating and orbital
time scales, significant radial variations  in $\tau\Omega$ 
would have implications for mean surface densities and fluxes.

{\bf Acknowledgments}:A.H. acknowledges a Horton fellowship from the UR Laboratory for  
Laser Energetics.
and E.B. acknowledges support from support from
NSF grant AST-0406799, NASA grant ATP04-0000-0016. We are grateful to
NORDITA  for support and hospitality during  the Turbulence and  
Dynamos workshop 2008.

\end{document}